# ARTICLE

## Unveiling the Role of Lewis Base Strength in Small-Molecule Passivation of Defect Perovskites

Yi-Chen Wu[a] and Hsien-Hsin Chou*[a]

Perovskite materials are highly promising for a range of optoelectronic applications including energy conversion technologies, owing to their high charge-carrier mobilities, adaptability of bandgap tuning, and exceptional light-harvesting capabilities. Yet, defects that arise during manufacturing often lead to performance limitations such as hindered efficiency and stability. This is primarily due to significant deviations in crystal geometry and band structure elements such as the Fermi level, work function, and density of states, compared to pristine perovskite. To mitigate these issues, this study explored the passivation of surface iodide-vacancy defect in perovskite using small-molecule Lewis bases, an approach aims to counteract these detrimental effects. Among the examined $N$-, $P$- and $O$-coordinated benzyl derivatives, those featuring a phosphonic acid group as a passivator for the undercoordinated Pb(II) sites demonstrated outstanding electronic structure properties. This was notably achieved by lowering the Fermi level, increasing the work function, and suppressing surface trap states. The effective restoration of electronic properties achieved by targeted small molecule passivation provides crucial insights into enhanced functionality and efficiency for defect perovskite materials.

## Introduction

Perovskite materials have emerged as an area of interest for their significant impact toward the development and enhancement of optoelectronic devices. Their unique properties, such as low exciton-binding energies, high charge-carrier mobility, broad absorption spectrum, and low fabrication costs make them ideal candidates for a variety of light-based technologies including solar cells, light-emitting diodes, lasers, and photodetectors.[1,2] Recently, the critical impact of trap states on the electronic properties of perovskite materials has been highlighted.[3–8] Although the unique defect tolerance, featuring shallow defects near the valence and conduction band edges, differentiates perovskite from silicon-based materials,[9–12] these defect states can act as charge trapping centers, leading to trap-assisted non-radiative recombination[13] and structural decomposition. Such defects can significantly impair the photoluminescence quantum efficiency and the functionality of photovoltaic devices even at very low concentrations.[5,14–23]

Recent advancements highlight the effective addressing of these challenges via post-synthetic surface passivation techniques. These strategies include either formation of the passivation layers via introduction of large organic cations or Lewis acid-base interactions between the perovskite material and passivators.[24–31] The latter approaches neutralize charges or induce chemical reactions that passivate defect sites.

Effective passivation has been demonstrated with a variety of materials including inorganic salts,[32,33] organic substrates,[9,21,25,34–36] and particularly the small organic molecules featuring functional groups like carbonyls[36–41] and phosphorus-containing groups.[42,43] For example, it is reported that the fullerene derivative can be used as buffer layer when exploited to the surface to release interface stress and therefore leading to suppressed interface defects.[44] Passivation of the defected perovskite using phenethylammonium iodide (PEAI) can effectively eliminate surface halogen deficiency to promote the efficiency of the perovskite solar cells to exceeding 23%.[14] After passivated with thiourea[45] or caprolactam,[46] the defect density of perovskites can be reduced by interaction of lone-pair electrons on coordinated atom of the small molecule with $Pb^{2+}$ at the surface and the grain boundary. A small-molecule dyad composed of phosphonate and phosphine oxide (PE-TPPO) has been employed as an additive in perovskite light-emitting diodes (PeLEDs), resulting in a maximum external quantum efficiency (EQE) of 25.1%.[47] The latter improvement, featuring a current efficiency of 100.5 cd $A^{-1}$ and a power efficiency of 98.7 lm $W^{-1}$, is attributed to both defect passivation and reduced non-radiative energy loss within the perovskite layer.

While extensive research has been devoted to the passivation of perovskite materials using small molecules, there remains a gap in the systematic understanding of how the electronic properties of defected perovskites are altered post-passivation. A recent study examines how P=O bonds in 2-aminoethylphosphonic acid (PA), S=O bonds in hypotaurine (SA), and C=O bonds in β-alanine (CA) influence the passivation performance and photostability of perovskite solar cells.[48] The findings emphasize the crucial role of Lewis basicity in the donor groups of passivation reagents, concluding that stronger localized negative charges on the oxygen atom lead to improved passivation. Using tribenzylphosphine oxide (TBPO) as a passivation agent results in considerably lower surface energy compared to triphenylphosphine oxide (TPPO), owing to the more effective P=O→Pb coordination of TBPO.[49] This leads to reduced surface activity of the perovskite towards the adsorption of water and other detrimental molecules.

[a] Department of Applied Chemistry, Providence University, Taichung 43301, Taiwan.

It is evident that well-established molecular designs for passivation strategies are crucial for the successful improvement of device performance. However, existing studies provide limited insight into the design principles of passivation molecules, highlighting a challenge in comprehensively understanding how specific functional groups within the molecules trigger effective passivation. In this article, we investigate the passivation of defects on the less stable 001 surface of methylammonium lead halide (MAPbI₃) perovskite materials using small-molecule Lewis bases. We aim to elucidate the underlying mechanisms that contribute to improved electronic performance in perovskite materials compromised by defects.

## Computational Details

First-principles calculations were performed using density functional theory (DFT) with the CASTEP program suite. The exchange–correlation interactions were described using the Perdew–Burke Ernzerhof (PBE) generalized gradient approximation, suitable for electronic calculations and geometry optimization. Ultra-soft pseudopotentials were utilized to describe the ionic cores and core-valence interactions, and the plane wave kinetic energy cut-off was set at 500 eV. The tetragonal MAPbI₃ perovskite material, featuring a 001 surface facet orientation, was modelled in a $2 \times 2 \times 1$ supercell containing 8 formula units ($8.56 \times 8.84 \times 44.3$ Å³). The Brillouin zone sampling was also set to $2 \times 2 \times 1$ for this model. To prevent interactions between periodic images, a vacuum region of approximately 25 Å was implemented for each model. Convergence criteria for energy and force were set at $10^{-5}$ Hartree and 0.002 eV/Å, respectively, with atomic positions fully relaxed to optimize the model. The maximum displacement and self-consistent field (SCF) convergence values were set to 0.005 Å and $10^{-6}$ Hartree, respectively. To effectively demonstrate defect passivation at unstable interfaces, the model was designed with a termination at the commonly used, yet less stable 001 surface, employing a PbI₂ layer instead of a MAI layer. A single iodide vacancy in MAPbI₃ was introduced by removing the 1-position on the 001 surface layer, as shown in Fig. 1. This study examines the defective structures with single vacancies in both cationic and neutral states. For passivated perovskite cases, small-molecule benzylic derivatives were coordinated on undercoordinated Pb at the axial and equatorial sites.

## Results and Discussion

To efficiently extract essential information while minimizing computational efforts, the model employs a $2 \times 2 \times 1$ slab configuration within the Brillouin zone. Initially, a pristine tetragonal MAPbI₃ crystal characterized by a 001 facet comprising eight formula units was developed to build models of defective perovskite. An iodine vacancy was introduced at the 1-position on the surface layer (Layer A) to create the cationic perovskite, as demonstrated in Fig. 1. All models incorporated a vacuum region of approximately 25 Å to facilitate the binding of small molecules onto the defective surface. As illustrated in Fig. 2a-d, surface passivation was achieved by coordinating a small-molecule Lewis base with the under-coordinated lead(II) center. This process resulted in the formation of cationic perovskite-small molecule adduct, abbreviated as PSK-sm. The Lewis bases examined in this study include phenyl and benzyl derivatives with nitrogen- or phosphorus-based donor groups, encompassing cyanide (PhCN), imine (PhCH=NH), primary amine (BzNH₂), phosphines (BzPR₂; R = H, Me, and Cl), phosphine oxide (BzP(=O)Me₂), and phosphonic acid (BzP(=O)(OH)₂), where Bz denotes to benzyl group (for the abbreviation of each molecules, see Scheme 1). In the case of **POOH2**, both oxygen atoms in the P=O and P-OH bonds are proposed to coordinate with Pb. These are identified as **POOH2-PO** and **POOH2-OH**, respectively. Additionally, we developed a series of neutral PSK-sm models corresponding to their cationic counterparts. For protic small molecules including **=NH**, **NH2**, **PH2**, **POOH2-PO**, and **POOH2-OH**, deprotonation from the donor groups results in the formation of neutral PSK-sm models (see Fig. 2e). In these scenarios, the local cationic framework in defect perovskite can thermodynamically release HI upon reaction with coordinated protic molecules, yielding neutral PSK-sm derivatives. In the case of **PCl2**, removing a chloride ion similarly results in a neutral PSK-PCl model. This represents the passivation of benzyldichlorophosphane on an iodide-defected perovskite surface, potentially leading to the release of iodine monochloride (ICl) and the formation of a PhCH₂PCl-passivated perovskite, possibly under oxidation or other specific conditions. Furthermore, the bonding of small molecules onto under-coordinated Pb was accomplished in two directions including axial- (6-position) and equatorial-sites (1-

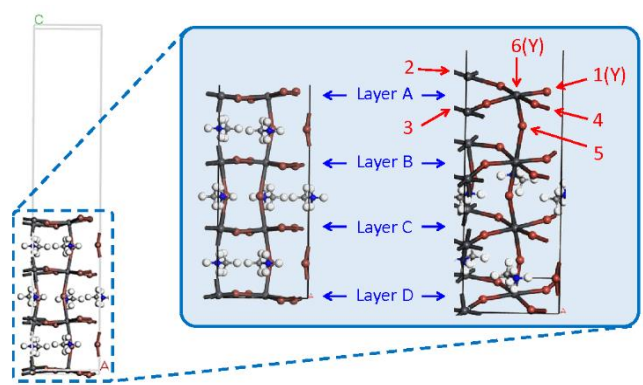

**Fig. 1.** Layer and atom indexes for the crystal structure of pristine perovskite CH₃NH₃PbI₃ with a 001 surface. The methyl ammonium cations within layers A and B were omitted for clarity.

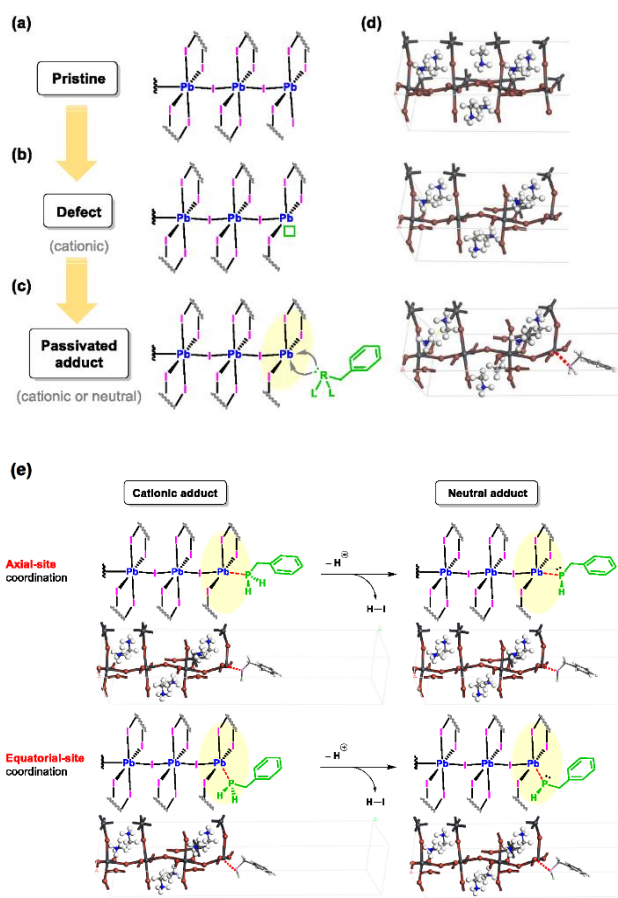

Fig. 2. Building of the defected perovskite models. Schematic drawing of the crystal structures for (a) the pristine, (b) the iodide-defected, and (c) the small-molecule bonding perovskite. (d) the optimized geometries for perovskites at stages (a) to (c). (e) Schematic representation for the definition of perovskite bonding sites as cationic adduct and generation of the neutral adduct. The green rectangle and the red dashed line indicate the site of coordination unsaturation and the bonding of small molecule with the coordination-unsaturated Pb atom, respectively.

position), representing the epical direction in line with that of interlayer Pb-I-Pb bonds and the position of iodide vacancy at the surface layer, respectively. Fig. S1 summarizes all the geometries for both cationic and neutral PSK-sm models in this study.

Fig. 3 and S2 illustrate the deviations in Pb-I and Pb-Y (Y = N, P or O) bond lengths ($L_{Pb-I/Y}$) within the surface layer A of each cationic PSK-sm model, compared to the pristine perovskite, $L_0$ (3.12 Å for the A-1 bond; 3.14-3.24 Å for A-2 to A-5 bonds). A notable characteristic of the defect perovskite in the absence of small molecule is the lattice shrinkage in terms of reduced Pb-I bond lengths at surface layer in accompanied with the shortened interlayer Pb-Pb distances (refer to Table S1 and S2 for details).[50] For part of the Pb-I bonds, the addition of a small molecule slightly mitigates this contraction, regardless of whether the models are in cationic/neutral forms or ax-/eq-site passivation is used. Furthermore, we observed notably extended Pb-P and Pb-O bond distances in both the axial and equatorial configurations of various molecules such as **PH2** (ax: 2.90 Å; eq: 2.94 Å), **PCl2** (ax: 2.98 Å ; eq: 3.21 Å), **PMe2** (eq: 2.87 Å), and **POOH2-OH** (ax: 2.79 Å; eq: 3.02 Å) in layer A.

Consequently, the slight deviations in the surface Pb-I(Y) bonds might lead to geometries that more closely resemble those of pristine perovskite.

Fig. 4a summarizes the calculated binding energy ($E_b$) for each PSK-sm model. Fig. S3 shows the potential energy surfaces corresponding to the approach of each small molecule towards the defect perovskite along the z-axis. The calculation is based on the energy difference between the small molecule-perovskite adduct and each isolated entity, given by $E_b = E_{PSK-sm}$ − ($E_{PSK} + E_{sm}$). Here $E_{PSK-sm}$, $E_{PSK}$, and $E_{sm}$ represent the calculated energies for PSK-sm models, the defect perovskite, and the passivating small molecules, respectively. The results clearly reveal several key insights: (1) $N$-based Lewis donors tend to have lower binding energies compared to $P$-based ones, as indicated by the ranges 0.40−0.73 eV versus 2.58−4.19 eV, respectively, for cationic system. This discrepancy in binding energies can largely be attributed to the differences in orbital interactions with the under-coordinated Pb sites. Phosphorus-based passivating molecules are capable of forming more robust covalent bonds, due to better orbital overlap with Pb, resulting in higher binding energies. In contrast, nitrogen-based molecules provide less effective stabilization of the complex, leading to lower binding energies due to their weaker interactions; (2) there is no significant difference in binding energies between axial (ax-) and equatorial (eq-) site adducts within each PSK-sm model. This consistency can be attributed to the constrained space near the surface binding sites and the fixed positions of existing binding iodides within the structurally distorted penta-coordinated complexes; (3) cationic adducts generally exhibit smaller binding energies than their corresponding neutral derivatives, illustrated by **PH2** (cationic: 2.58−2.61 eV; neutral: 4.29−4.32 eV), **POOH2-PO** (cationic: 3.00−3.01 eV; neutral: 6.23−6.46 eV), and **POOH2-OH** (cationic: 2.58−2.98 eV; neutral: 4.81−4.85 eV) systems.

More interesting findings were found for electronic properties between the cationic and neutral system including work function and band structure. To ensure more accurate estimation of work function, the isolated iodide/hydrogen iodide present in the constrained void space of the cationic/neutral system was excluded to eliminate any potential interference. This is illustrated in Fig. S4 where electrostatic potential probed for both cationic and neutral PSK-sm adduct models were compared. A notable distinction between pristine and defect perovskites lies in their work functions, which are 5.82 eV and 3.59 eV, respectively, as shown in Fig. 4b. The work function is defined as the energy difference between the Fermi level ($E_F$) and the vacuum level. Consequently, the substantially reduced work function in passivator-free defect perovskite indicates a marked elevation of $E_F$. This observation aligns with independent findings from Cai's group,[50] which examined iodide-vacancy defects in two-dimensional monolayer-PbI$_2$ within a 6 × 6 × 1 supercell. They noted a decrease in work functions from 5.91 eV in pristine PbI$_2$ to 5.58 eV with a single vacancy and further to 4.60 eV with concentrated vacancies. Additionally, our studies on models of passivated perovskites reveal that cationic adduct models (2.86-3.59 eV) have smaller work functions compared to their neutral counterparts (4.69-5.67 eV). This phenomenon has been experimentally validated,[51] where the work function of mixed halide perovskites can be reduced through the use of protonated phenformin, achieved via passivation at both the top surface and grain boundaries, or solely at the top surface. In our analysis, we observed that the difference in work functions between both model types is distinctly marked by a threshold of approximately 4.14 eV, which is virtually centered between the two. While most cationic PSK-sm models exhibit slightly smaller work functions than passivator-free perovskite, the cases of POMe$_2$ (ax: 2.86 eV; eq: 3.06 eV) represent a lower extreme among the rest. The two bulky methyl substituents at the sterically congested binding sites near the iodide vacancy may have an additional influence on the electronic properties of the system. Another notable difference lies in the neutral type phosphonate adducts, abbreviated as **PO2OH-PO** and **PO2OH-OH**, after releasing HI. These adducts exhibit larger work functions of 5.52 eV and 5.67 eV for the axial and equatorial sites of **PO2OH-PO**, respectively, and 5.64 eV and 5.52 eV for the axial and equatorial sites of **PO2OH-OH**, respectively. These values exceed those of other derivatives (4.69–5.30 eV), and approach that of pristine materials (5.82 eV). This phenomenon suggests that the coordination of phosphonate to the under-coordinated Pb, via unique P–O→Pb bonding[47,49,52–54] will significantly contribute to the stabilization of the Fermi level and the restoration of bulk crystal properties.

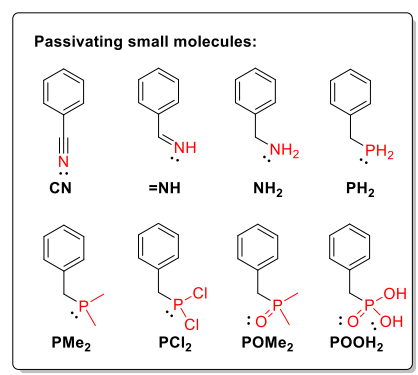

**Scheme 1.** Molecular structures for the small molecules used in this study. The annotated lone pairs indicate atoms to be coordinated to the perovskite surface.

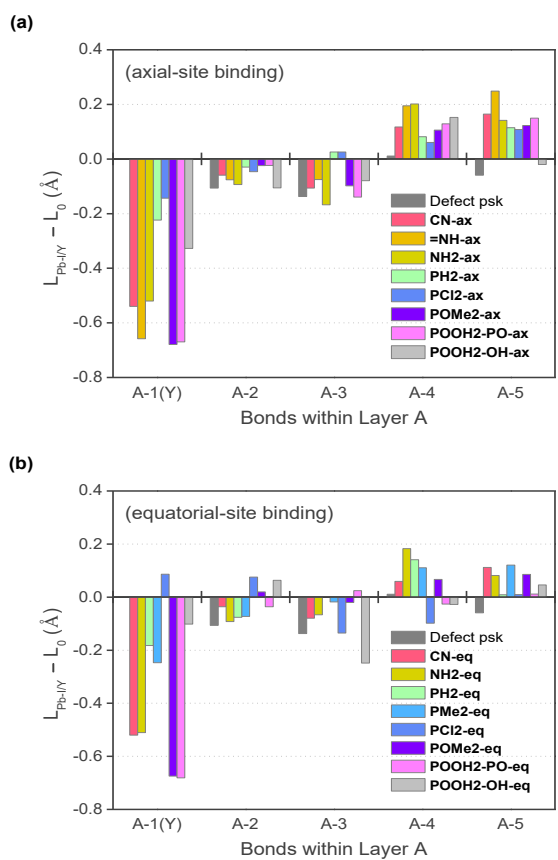

**Fig. 3.** The deviation in bond lengths relative to the pristine perovskite for bonds within the surface layer. (a) the cationic axial-coordinated PSK-sm models, and (b) the cationic equatorial-coordinated PSK-sm models. The Pb atom and the remaining atoms in the surface layer (layer A) are denoted as A, 1, 2, and so forth for clarify.

Further comparison between the density of states (DOS) for the pristine and vacancy-defect perovskite at 001 surface was depicted in the upper section of Fig. 5a. The pristine case exhibits a bandgap of approximately 1.2 eV with the Fermi level ($E_F$) located at the edge of the valence band. This finding shows a minor discrepancy from the results reported by Targhi et al.[55] who observed a bandgap of 1.64 eV using 6 × 6 × 6 k-point grid in the Brillouin zone. The difference may also be influenced by their use of different orientations (111 and 100). Furthermore, removing an iodide from the surface PbI$_2$ layer into the vacuum for cationic PSK-sm systems in accompanied with partial DOS contributed from Pb (6s, 6p), I (5p), and the small molecules. The interaction between the under-coordinated Pb and the Lewis donor results in a significant narrowing of the bandgap. This occurs primarily due to a substantial downward extension of the conduction band edge. Meanwhile, the introduction of additional states from Lewis donor results in a considerable variation in the positions of the corresponding Fermi levels. Class A cationic PSK-sm adducts, including **CN-eq**, **PCl2-ax/eq**, **POOH2-PO-eq**, and **POOH2-OH-ax/eq**, exhibit a notable shift of the Fermi level toward the valence band, while approaching TS. This considerable shift makes the corresponding band structure closely resemble that of the neutral pristine PSK. Similarly, Class B (sm = **NH2-ax/eq**, **PMe2-eq**, **POOH2-PO-ax**) and Class C (sm = **CN-ax**, **=NH-ax**, **PH2-ax/eq**, **POMe2-ax/eq**) cationic PSK-sm adduct systems display moderate and smaller Fermi level shifts, respectively, each accompanied by a correspondingly narrower bandgap. Eventually, both Class B and C systems show the Fermi level positioned within the lower edge of the conduction band, which may be attributed to a significant downward extension of TS. Interestingly, the combined features including extended trap states and bandgap narrowing[57] become prominent after passivation. This effect is largely ascribed to the interaction between undercoordinated Pb 6s and 6p orbitals and specific Lewis donors from the small molecule, as depicted in the partial DOS shown in Fig. 6b. Additionally, the presence of an iodide vacancy results in the formation of trap states, which range from shallow to somewhat deep levels alongside the valence band, affecting iodide 5p orbitals. This phenomenon is accompanied by varied shifts in the Fermi level, as illustrated in the partial DOS in Fig. 6a. Fig. 7 summarizes the observations discussed above, demonstrating the varied interactions between small molecules and surface Pb (6s, 6p) orbitals. Phosphines are generally accepted as stronger Lewis bases compared to amines, owing to their greater polarizability and stronger nucleophilicity. Conversely, the oxygen atom on P=O group of phosphine oxide and phosphonic acid act as weaker Lewis bases compared to the former two, due to their resonance-stabilized structures and electronegative oxygen atoms. Therefore, small molecules bearing the least basic Lewis donating groups, such as BzPCl₂ and BzP(=O)(OH)₂ in Class A system, form lower-energy antibonding orbitals that stabilize TS closer to the valence band edge, due to weaker interactions with Pb (6s, 6p). In contrast, stronger Lewis donating groups in Class C small molecules, like BzPH₂ and PhCN, which have a strong interaction with coordination-unsaturated Pb, leading to an elevated distribution of TS energy levels at the valence band edge and narrowed bandgap. As a result, the Class C PSK-sm system exhibits a Fermi level and DOS similar to those observed in the passivation-free case, including extrinsic shallow TS, as illustrated in the lower sections of both Figures 5a and 5b. This is evident in the dramatic differences between onset energies of valence band in Class A (-0.103 – 0.122 eV vs $E_F$) versus those in Class B (-0.668 – -0.370 eV vs $E_F$) and Class C (-0.982 – -0.763 eV vs $E_F$) system, shown in Fig. S5. Interestingly, our findings on the cationic PSK-sm system align with reported results on the space creates a vacancy defect, narrowing the bandgap due to the formation of trap states (TS) at both the valence and conduction band edges, as shown in the middle section of Fig. 5a.[56,57] This also causes elevation of $E_F$ by about 1.7 eV. Similarly, experimental evidence reported by Cho et al. shows that Cs-doped mixed cation $(FA_{0.9}MA_{0.1})Pb(I_xBr_{1-x})$ exhibits a shift of the $E_F$ closer to the conduction band by 0.17 eV when the molarity of the solution containing PbI₂ is increased from 1.1 to 1.5 M.[58] For a clearer comparison, annotation on the Figure with an arrow illustrates a shift of the entire DOS to better align with the pristine case. Fig. 5b and Fig. 6a-6c displays the total DOS anionic I-dimer defects on the PbI₂ surface, where a notable shift in the TS within the DOS gap toward the valence band was noted after passivation with ionic-type CsF.[59] This change is attributed to altered surface charge localization post-passivation, which either converts deep TS into shallower ones or eliminates TS altogether.[8] Furthermore, the emergence of geometries resembling those in the pristine case, following small-molecule passivation, aids in recovering from reduced bond distances. This allows for the establishment of a structure-property relationship based on extrinsic bond properties. The corresponding edge energies of the valence bands, which correlate with the strength of orbital interactions, are inversely related to the variations in the average bond lengths across the top two surface layers, namely layers A and B. As demonstrated in Fig. 8, a moderate correlation coefficient of 0.67 was obtained for the cationic PSK-sm systems. This is likely due to the difficulty in estimating edge energies, influenced by the presence of shallow traps and limited void space on the perovskite surface during the optimization process.

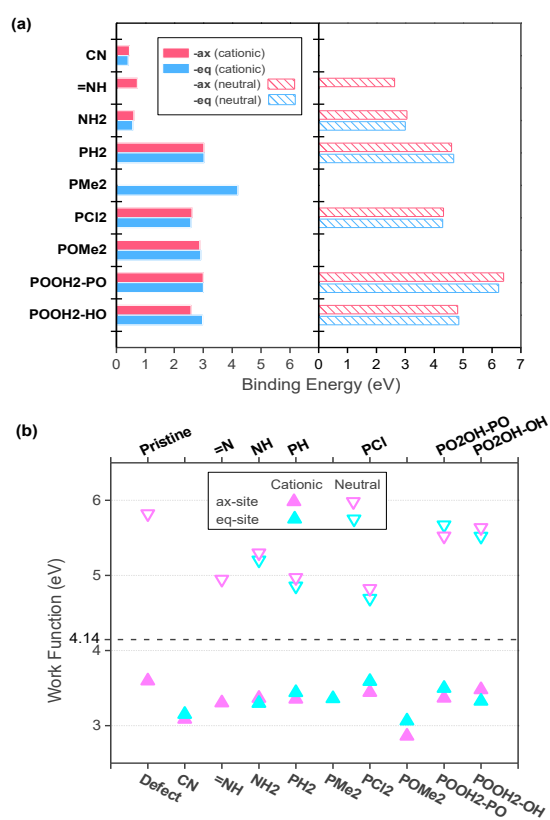

**Fig. 4.** (a) Binding energies calculated for bonding of the defected perovskite with small molecules. (b) Work functions for cationic (solid dots) and neutral (circles) PSK-sm adduct models.

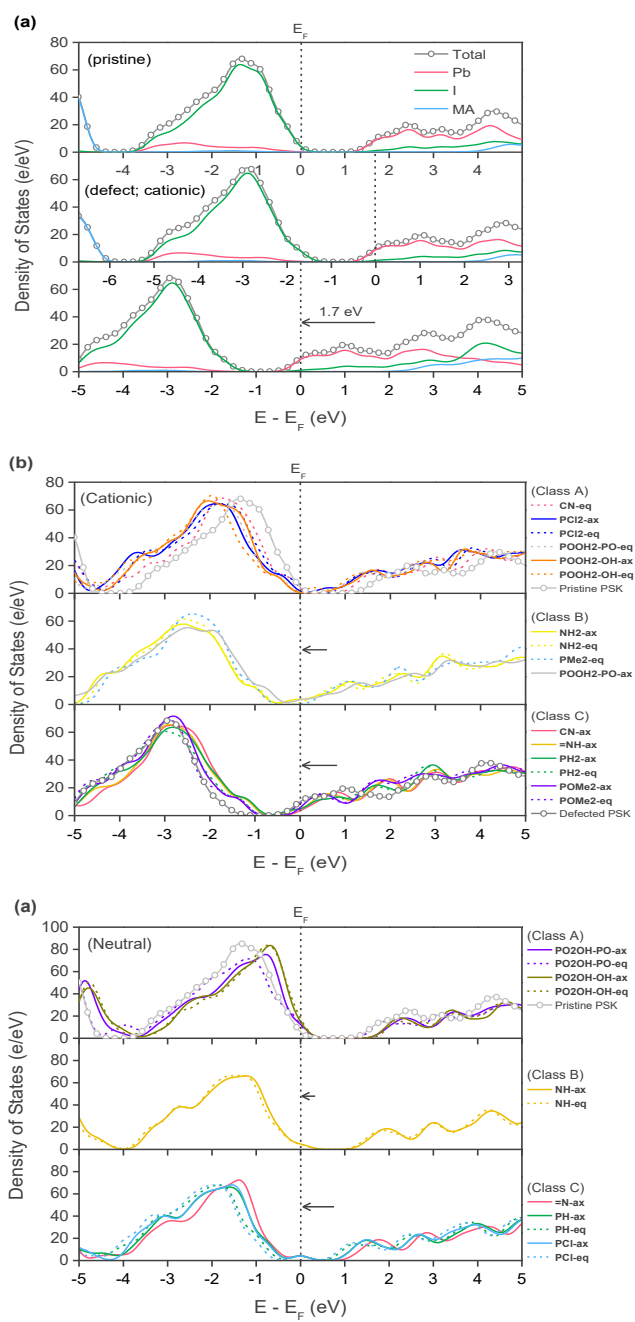

**Fig. 5.** Total density of states for both pristine and defect PSK, as well as passivated PSK-sm systems. Arrows highlight the approximate shifts in Fermi levels, depicted as dashed vertical lines for clear alignment.

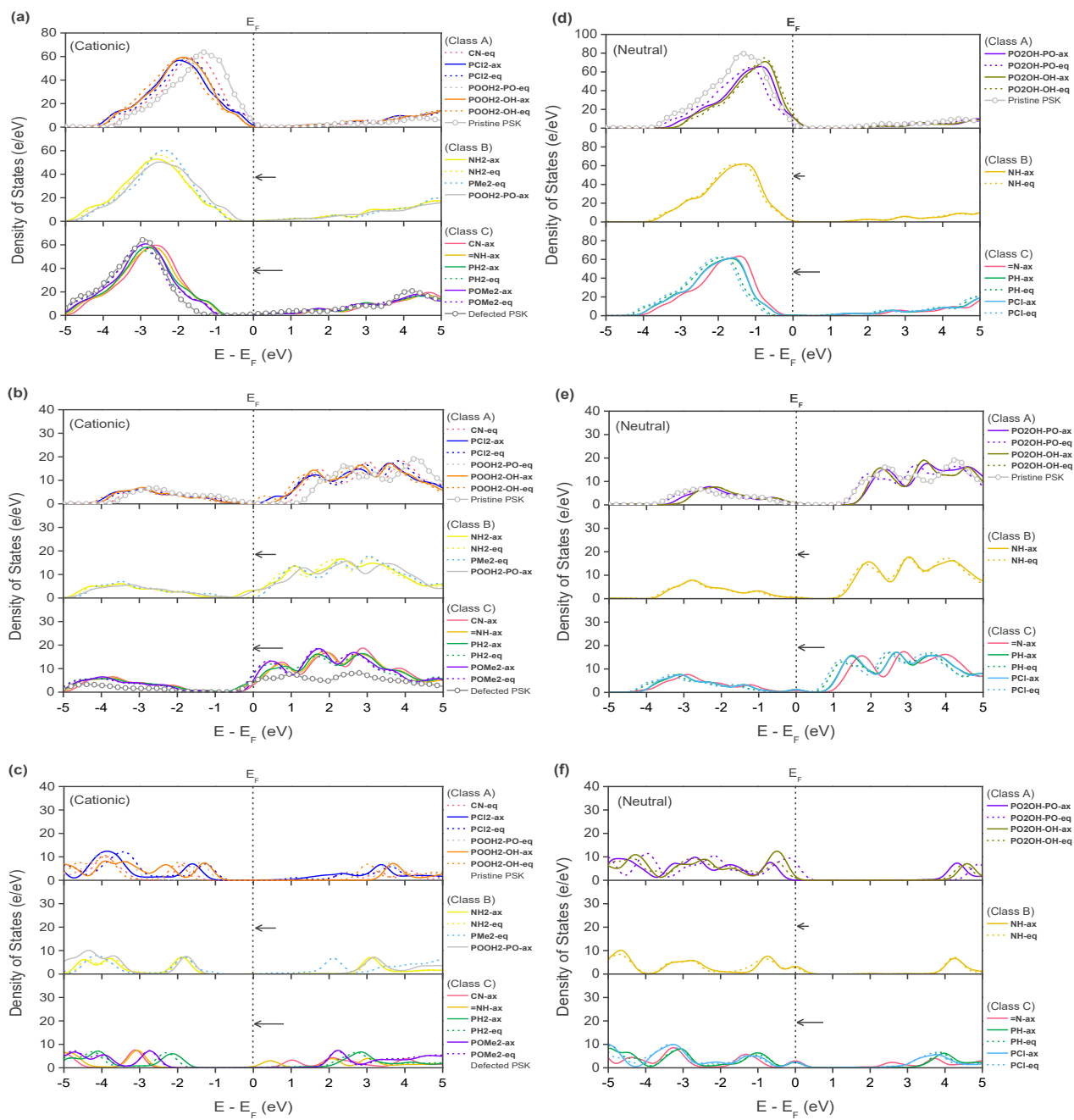

**Fig. 6.** Partial Density of States (PDOS) for cationic (panels a-c) and neutral (panels d-f) PSK-sm systems, detailing contributions from I (5p) orbitals in panels a and d, Pb (6s, 6p) orbitals in panels b and e, and small molecules in panels c and f. The cationic and neutral systems each have independent separate legends. Arrows mark the approximate shifts in Fermi levels, indicated by dashed vertical lines for precise alignment.

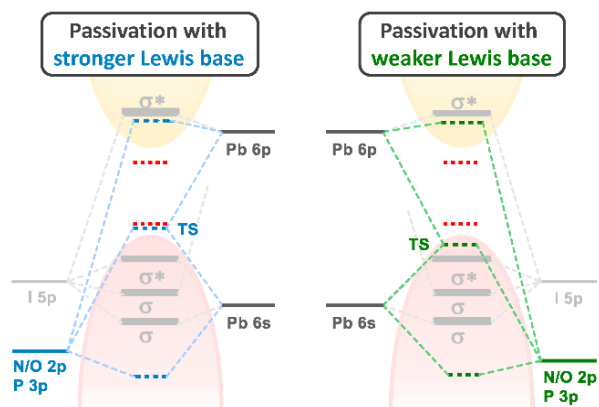

**Fig. 7.** Impact of Lewis base strength on trap state (TS) variations in PSK-sm systems, illustrated with dashed lines. Red dotted lines indicate TS in defect perovskite, while blue and green dashed lines represent TS influenced by stronger and weaker Lewis bases, respectively.

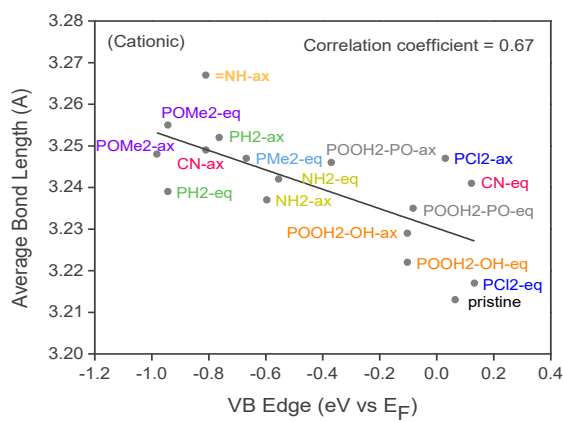

**Fig. 8.** Relationships between average bond length and valence band edge for cationic PSK-sm systems.

Finally, an interesting observation focuses on the neutral PSK-sm systems, which exhibit slight differences compared to their cationic counterparts. The resulting small molecules, following the release of a proton or a chloride from the cationic adduct, are classified into Class A (for **PO2OH-PO**, sm = BzP(=O)$_2$(OH)), Class B (for **NH**, sm = BzNH$^-$), and Class C (for **=N**, sm = PhCH=N$^-$; for **PH**, sm = BzPH$^-$; for **PCl**, sm = BzPCl$^-$) neutral PSK-sm systems (see Fig. 5c). Unlike the cationic PSK-sm systems, the overall density of states for each class in neutral system exhibits a similar energy distribution profile near the bandgap. Consequently, there is less significant shift in the Fermi levels across the three classes of PSK-sm systems. However, the positioning of TS near the valence band edge is a critical factor that leads to dramatically different outcomes. The trap states are primarily contributed by the essentially anionic small-molecule ligands, with a minor contribution from Pb (6s, 6p) orbitals, as depicted in Fig. 6d-6f. Similar to the cationic case, Class A systems, such as **PO2OH-PO** and **PO2OH-OH**, show shallow TS at the valence band edge. In this scenario, TS become closer to the valence band edge, which is beneficial for suppressing Shockley-Read-Hall recombination and extending the exciton lifetime.[9,10] Conversely, deep traps located within the bandgap, as observed in Class C systems, are considered detrimental to the efficiency of perovskite-based devices such as solar cells and light-emitting diodes.[9,60] Unlike the cationic case, these deep traps are less likely to participate in conduction or valence processes, rendering them less influential on the materials' electronic properties. This also brings certain difficulties in estimating VB edge. As shown in Fig. S6, only the Class A system possesses a similar negative relationship between the VB edge and average bond lengths.

## Conclusions

In conclusion, this study explores small-molecule passivation on archetype iodide-vacancy-defected MAPbI$_3$ perovskites using density functional theory (DFT) calculations. A comprehensive investigation into the influences of various small-molecule Lewis bases was conducted, with a focus on improving the crystal properties of defect perovskite. Such passivation slightly diminishes the contracted Pb-I bonds within the crystal structure, resulting in a geometry more closely resembling the pristine state. As evidenced by the calculated bandgaps, Fermi levels, work functions, and density of states, we demonstrate that both the geometrical and electronic properties of these perovskites can be effectively controlled through molecular engineering on small molecules, as established by a relationship between bonding distance and the edge energy of valence band. Particularly, passivation with benzyl phosphonic acid/phosphonate is highlighted for its potential ability to inhibit the formation of deep traps and lower the Fermi level. The optimized perovskite with repaired electronic structure can effectively reduce trap-mediated non-radiative recombination at surface defects, crucial for enhancing the separation and transport of charge carriers. This improvement is expected to significantly boost in performance for applications such as solar cells, PeLEDs, photodetectors, and lasers. The study establishes a foundation for the development of more robust and sustainable perovskite materials, addressing key challenges in optoelectronic applications.

## Author contributions

Y.-C. W. conducted the DFT calculations and analysed the data. All the authors discussed the results and commented on the manuscript.

## Conflicts of interest

The authors declare no conflicts of interest.

## Acknowledgements

The authors thank the financial support for this work from the National Science and Technology Council (NSTC) in Taiwan with Grant No. MOST 111-2113-M-126-002 and NSTC 112-2113-M-126-002. We thank to National Center for High-performance Computing (NCHC) of National Applied Research Laboratories (NARLabs) in Taiwan for providing computational and storage resources.

**Supporting Information**

**Unveiling the Role of Lewis Base Strength in Small-Molecule Passivation of Defect Perovskites**

*Yi-Chen Wu and Hsien-Hsin Chou\**

Department of Applied Chemistry, Providence University, Taichung 43301, Taiwan

E-mail: hhchou@pu.edu.tw

**Fig. S1.** Naming and optimized geometries for the perovskite-small molecule adducts.

**(a)**

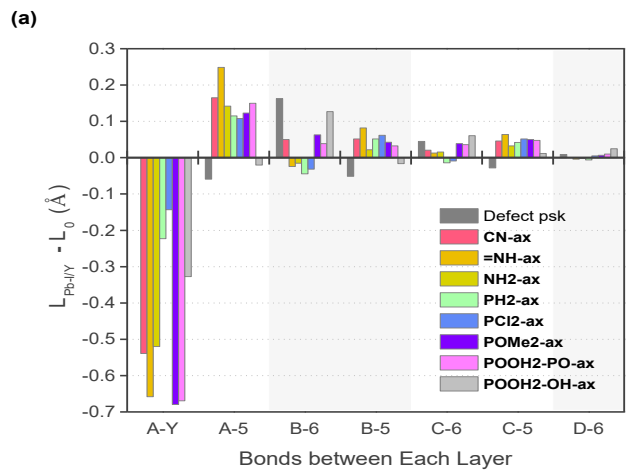

**(b)**

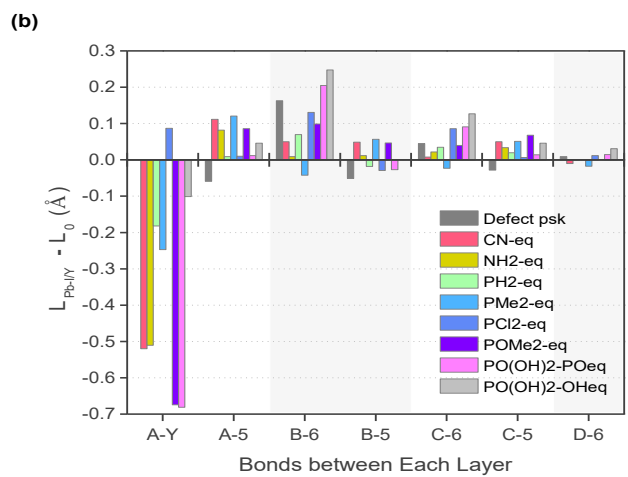

**Fig. S2.** The deviation in bond lengths relative to the pristine perovskite for bonds between each layer. (a) the axial-coordinated PSK-sm models, and (b) the equatorial-coordinated PSK-sm models. The Pb atom and the remaining atoms in the surface layer (layer A) are denoted as A, 1, 2, and so forth for clarify.

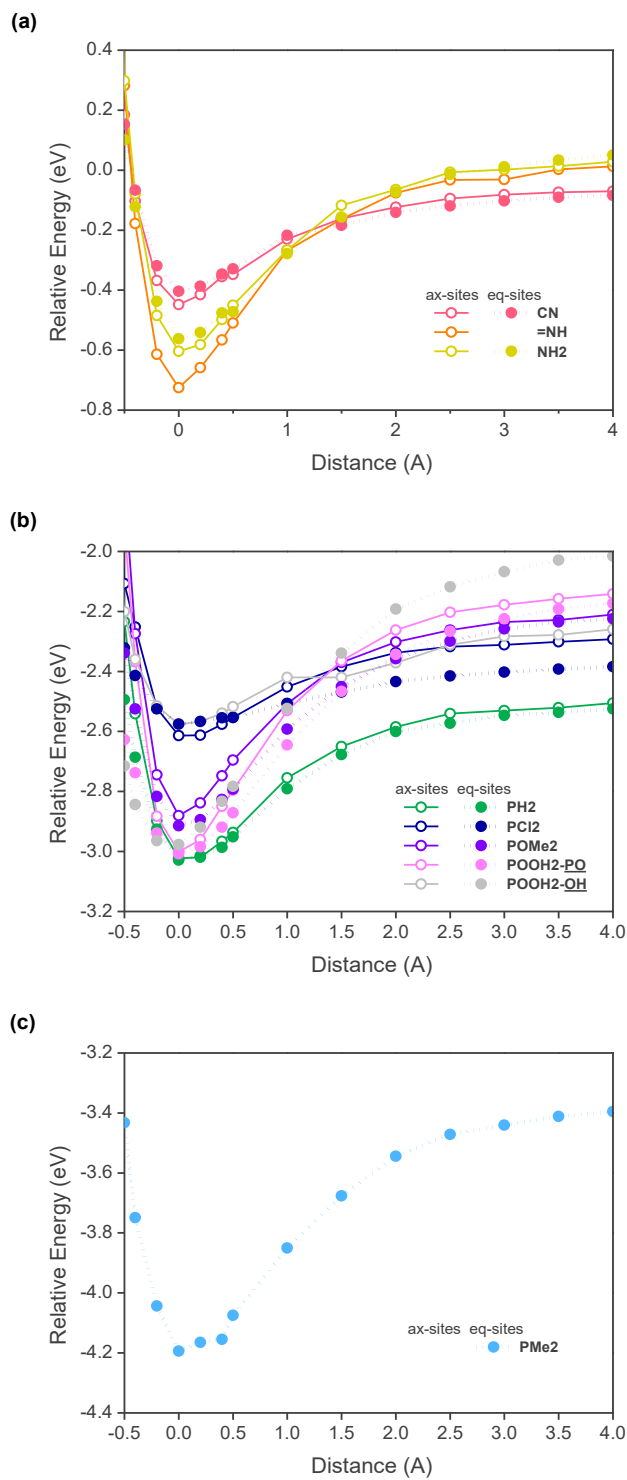

**Fig. S3.** Potential energy surface calculated for the interaction of defected perovskite with small molecules along the z-axis. (a) N-type donors, (b) P-type and P=O type donors, and (c) PMe2 donor.

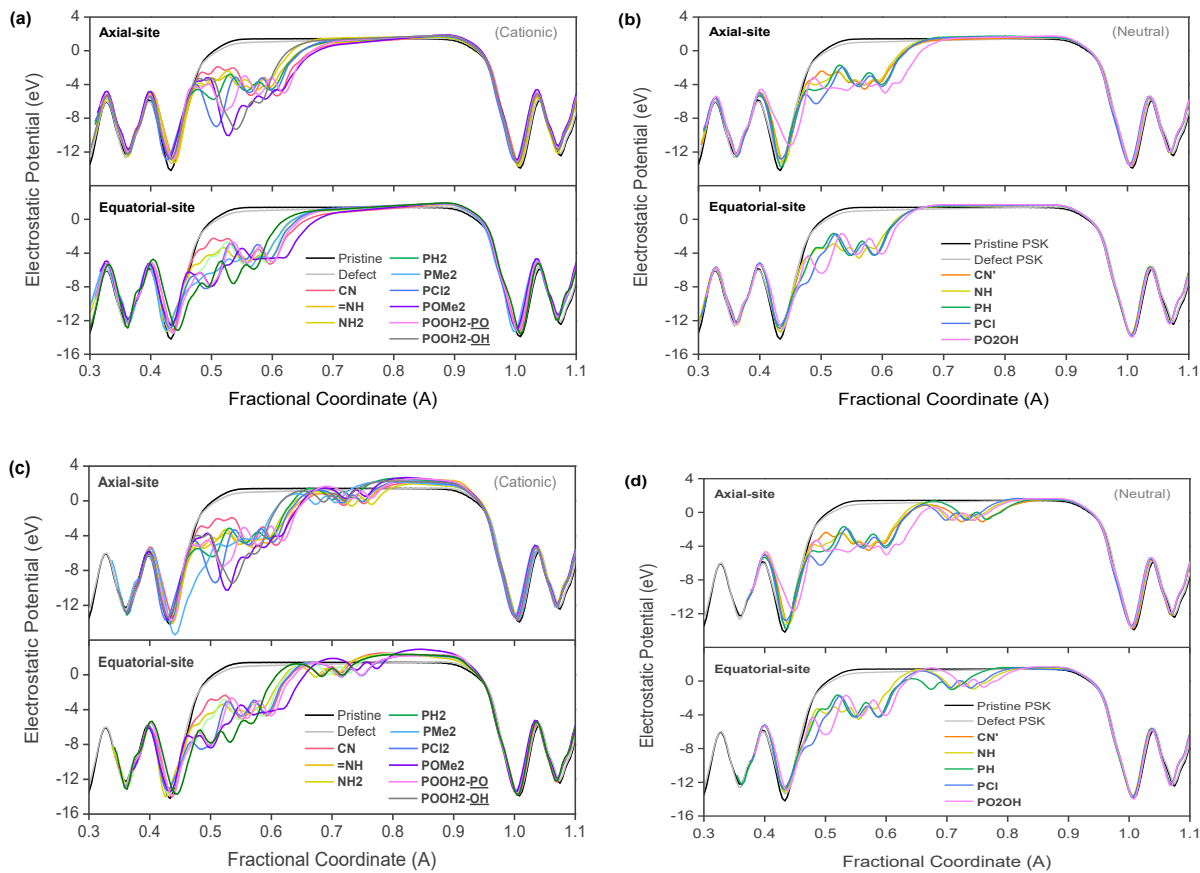

**Fig. S4.** Electrostatic potential probed for PSK-sm models with and without the inclusion of iodide/hydrogen iodide at void space. (a) cationic PSK-sm, (b) neutral PSK-sm, (c) cationic PSK-sm with I/HI, and (d) neutral PSK-sm with I/HI.

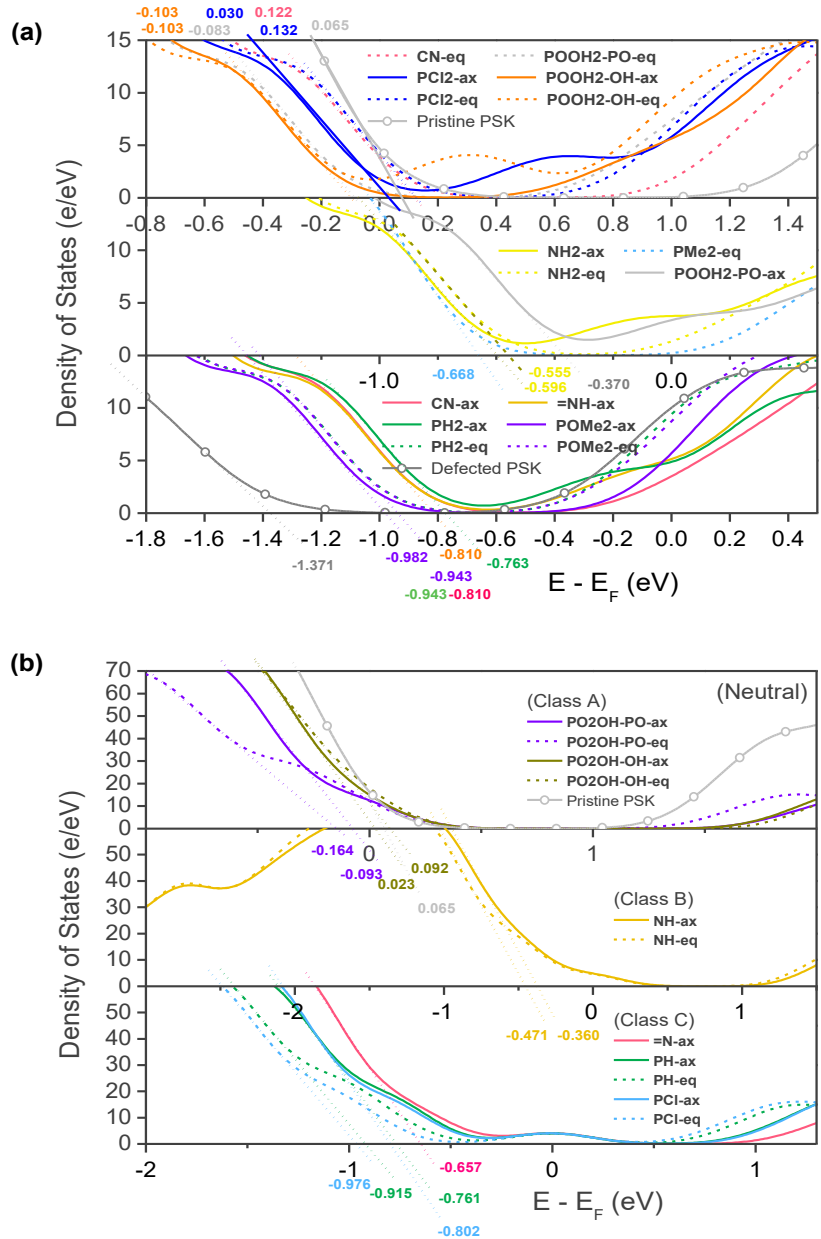

**Fig. S5.** Edge energies of the valence band (VB) in the density of states for three classes of passivated models within the (a) cationic and (b) neutral PSK-sm systems. For the neutral system, the bulk density of states is utilized to estimate VB edge since the trap states predominantly appear as deep traps within the bandgap.

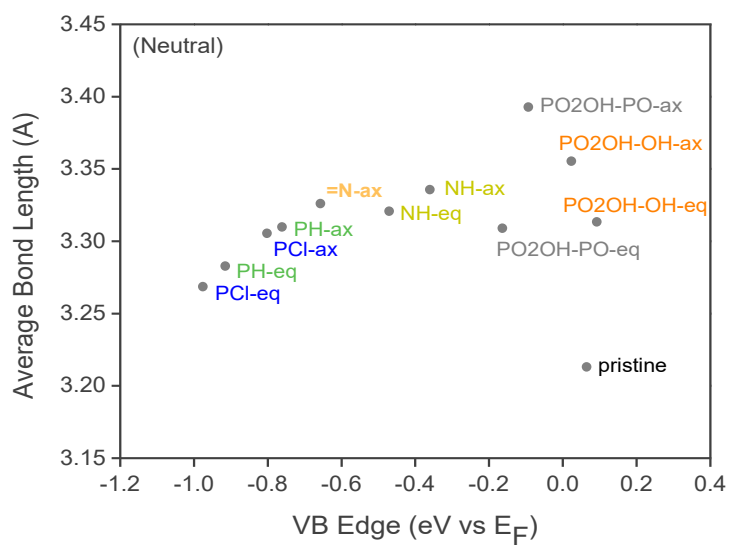

**Fig. S6.** Relationships between average bond length and valence band edge for neutral PSK-sm systems.

**Table S1.** Bond lengths in cationic and neutral PSK-sm system.

| | prinsitine | defect | CN-ax | =NH-ax | NH2-ax | PH2-ax | | PCl2-ax | POMe2-ax | POOH2-PO-ax | POOH2-OH-ax |
|---|---|---|---|---|---|---|---|---|---|---|---|
| Cationic PSK-sm system | | | | | | | | | | | |
| A-1/ A-Y | 3.120 | | 2.581 | 2.462 | 2.600 | 2.897 | | 2.977 | 2.441 | 2.451 | 2.793 |
| A-2 | 3.192 | | 3.133 | 3.116 | 3.099 | 3.163 | | 3.146 | 3.168 | 3.169 | 3.087 |
| A-3 | 3.320 | | 3.214 | 3.245 | 3.153 | 3.346 | | 3.346 | 3.222 | 3.181 | 3.241 |
| A-4 | 3.242 | | 3.360 | 3.437 | 3.444 | 3.324 | | 3.303 | 3.348 | 3.371 | 3.395 |
| A-5 | 3.135 | | 3.300 | 3.384 | 3.277 | 3.250 | | 3.243 | 3.258 | 3.285 | 3.115 |
| Averaged Layer A | 3.202 | | 3.252 | 3.296 | 3.243 | 3.271 | | 3.260 | 3.249 | 3.252 | 3.210 |
| B-1 | 3.224 | | 3.224 | 3.218 | 3.215 | 3.238 | | 3.243 | 3.229 | 3.228 | 3.231 |
| B-2 | 3.193 | | 3.198 | 3.202 | 3.209 | 3.226 | | 3.233 | 3.224 | 3.220 | 3.216 |
| B-3 | 3.218 | | 3.227 | 3.234 | 3.242 | 3.233 | | 3.216 | 3.230 | 3.227 | 3.225 |
| B-4 | 3.224 | | 3.230 | 3.226 | 3.221 | 3.208 | | 3.197 | 3.205 | 3.207 | 3.211 |
| B-5 | 3.233 | | 3.285 | 3.315 | 3.255 | 3.285 | | 3.295 | 3.276 | 3.266 | 3.217 |
| B-6 | 3.259 | | 3.309 | 3.235 | 3.244 | 3.215 | | 3.228 | 3.322 | 3.298 | 3.386 |
| Averaged Layer B | 3.225 | | 3.246 | 3.238 | 3.231 | 3.234 | | 3.235 | 3.248 | 3.241 | 3.248 |
| C-1 | 3.242 | | 3.262 | 3.263 | 3.263 | 3.260 | | 3.243 | 3.242 | 3.245 | 3.248 |
| C-2 | 3.217 | | 3.227 | 3.222 | 3.218 | 3.217 | | 3.203 | 3.216 | 3.217 | 3.216 |
| C-3 | 3.183 | | 3.184 | 3.184 | 3.183 | 3.196 | | 3.211 | 3.208 | 3.204 | 3.204 |
| C-4 | 3.198 | | 3.203 | 3.209 | 3.213 | 3.224 | | 3.237 | 3.210 | 3.210 | 3.210 |
| C-5 | 3.259 | | 3.305 | 3.323 | 3.292 | 3.301 | | 3.311 | 3.309 | 3.307 | 3.271 |
| C-6 | 3.218 | | 3.239 | 3.231 | 3.234 | 3.204 | | 3.209 | 3.257 | 3.254 | 3.279 |
| Averaged Layer C | 3.220 | | 3.237 | 3.239 | 3.234 | 3.234 | | 3.236 | 3.240 | 3.240 | 3.238 |
| D-1 | 3.138 | | 3.129 | 3.139 | 3.132 | 3.133 | | 3.135 | 3.123 | 3.123 | 3.114 |
| D-2 | 3.188 | | 3.180 | 3.182 | 3.181 | 3.187 | | 3.195 | 3.211 | 3.201 | 3.194 |
| D-3 | 3.297 | | 3.319 | 3.313 | 3.317 | 3.324 | | 3.323 | 3.344 | 3.337 | 3.346 |

| | | | CN-eq | | NH2-eq | PH2-eq | PMe2-eq | PCl2-eq | POMe2-eq | POOH2-PO-eq | POOH2-OH-eq |
|---|---|---|---|---|---|---|---|---|---|---|---|
| D-4 | 3.232 | | 3.246 | 3.251 | 3.245 | 3.246 | | 3.237 | 3.222 | 3.229 | 3.230 |
| D-6 | 3.127 | | 3.127 | 3.123 | 3.125 | 3.121 | | 3.132 | 3.134 | 3.137 | 3.152 |
| Averaged Layer D | 3.196 | | 3.200 | 3.202 | 3.200 | 3.202 | | 3.204 | 3.207 | 3.205 | 3.207 |
| A-B | 6.310 | | 6.482 | 6.453 | 6.310 | 6.326 | | 6.330 | 6.395 | 6.379 | 6.303 |
| B-C | 6.355 | | 6.433 | 6.452 | 6.390 | 6.386 | | 6.410 | 6.429 | 6.418 | 6.397 |
| C-D | 6.305 | | 6.352 | 6.364 | 6.331 | 6.343 | | 6.367 | 6.351 | 6.352 | 6.336 |
| | | | **CN-eq** | | **NH2-eq** | **PH2-eq** | **PMe2-eq** | **PCl2-eq** | **POMe2-eq** | **POOH2-PO-eq** | **POOH2-OH-eq** |
| A-1(Y) | | | 2.600 | | 2.610 | 2.938 | 2.873 | 3.207 | 2.446 | 2.439 | 3.019 |
| A-2 | | | 3.156 | | 3.100 | 3.116 | 3.120 | 3.268 | 3.212 | 3.156 | 3.256 |
| A-3 | | | 3.241 | | 3.254 | 3.320 | 3.302 | 3.185 | 3.300 | 3.345 | 3.072 |
| A-4 | | | 3.302 | | 3.425 | 3.383 | 3.353 | 3.144 | 3.309 | 3.216 | 3.215 |
| A-5 | | | 3.247 | | 3.217 | 3.144 | 3.256 | 3.145 | 3.221 | 3.147 | 3.181 |
| Averaged Layer A | | | 3.237 | | 3.249 | 3.241 | 3.258 | 3.186 | 3.261 | 3.216 | 3.181 |
| B-1 | | | 3.216 | | 3.225 | 3.221 | 3.238 | 3.227 | 3.218 | 3.210 | 3.215 |
| B-2 | | | 3.207 | | 3.215 | 3.214 | 3.235 | 3.211 | 3.217 | 3.203 | 3.198 |
| B-3 | | | 3.239 | | 3.239 | 3.233 | 3.241 | 3.241 | 3.226 | 3.234 | 3.220 |
| B-4 | | | 3.224 | | 3.217 | 3.216 | 3.202 | 3.219 | 3.194 | 3.208 | 3.208 |
| B-5 | | | 3.282 | | 3.245 | 3.215 | 3.290 | 3.204 | 3.280 | 3.206 | 3.233 |
| B-6 | | | 3.309 | | 3.268 | 3.329 | 3.217 | 3.390 | 3.358 | 3.464 | 3.507 |
| Averaged Layer B | | | 3.246 | | 3.235 | 3.238 | 3.237 | 3.249 | 3.249 | 3.254 | 3.264 |
| C-1 | | | 3.266 | | 3.258 | 3.253 | 3.237 | 3.242 | 3.238 | 3.245 | 3.247 |
| C-2 | | | 3.227 | | 3.216 | 3.212 | 3.201 | 3.217 | 3.216 | 3.221 | 3.228 |
| C-3 | | | 3.184 | | 3.189 | 3.192 | 3.209 | 3.198 | 3.206 | 3.195 | 3.190 |
| C-4 | | | 3.208 | | 3.218 | 3.218 | 3.231 | 3.211 | 3.211 | 3.200 | 3.191 |
| C-5 | | | 3.309 | | 3.293 | 3.279 | 3.310 | 3.266 | 3.327 | 3.273 | 3.305 |
| C-6 | | | 3.226 | | 3.240 | 3.253 | 3.195 | 3.304 | 3.258 | 3.309 | 3.345 |
| Averaged Layer C | | | 3.237 | | 3.236 | 3.235 | 3.231 | 3.240 | 3.243 | 3.241 | 3.251 |

| | prinsitine | defect | | =N-ax | NH-ax | PH-ax | | PCl-ax | | PO2OH-PO-ax | PO2OH-OH-ax |
|---|---|---|---|---|---|---|---|---|---|---|---|
| D-1 | | | 3.132 | | 3.134 | 3.130 | 3.146 | 3.127 | 3.130 | 3.116 | 3.107 |
| D-2 | | | 3.177 | | 3.187 | 3.184 | 3.192 | 3.181 | 3.208 | 3.197 | 3.203 |
| D-3 | | | 3.318 | | 3.313 | 3.316 | 3.306 | 3.313 | 3.331 | 3.334 | 3.360 |
| D-4 | | | 3.250 | | 3.239 | 3.241 | 3.239 | 3.239 | 3.228 | 3.231 | 3.233 |
| D-6 | | | 3.117 | | 3.128 | 3.127 | 3.110 | 3.139 | 3.127 | 3.142 | 3.158 |
| Averaged Layer D | | | 3.199 | | 3.200 | 3.200 | 3.199 | 3.200 | 3.205 | 3.204 | 3.212 |
| A-B | | | 6.449 | | 6.295 | 6.629 | 6.287 | 6.289 | 6.468 | 6.449 | 6.492 |
| B-C | | | 6.414 | | 6.387 | 6.373 | 6.380 | 6.410 | 6.442 | 6.418 | 6.488 |
| C-D | | | 6.343 | | 6.340 | 6.327 | 6.340 | 6.330 | 6.348 | 6.323 | 6.371 |
| Neutral PSK-sm system | | | | | | | | | | | |
| | prinsitine | defect | | =N-ax | NH-ax | PH-ax | | PCl-ax | | PO2OH-PO-ax | PO2OH-OH-ax |
| A-1(Y) | | | | 2.189 | 2.172 | 2.711 | | 2.726 | | 2.325 | 2.161 |
| A-2 | | 3.086 | | 3.141 | 3.264 | 3.259 | | 3.433 | | 3.292 | 3.176 |
| A-3 | | 3.183 | | 3.308 | 3.295 | 3.23 | | 3.189 | | 3.422 | 3.238 |
| A-4 | | 3.253 | | 3.489 | 3.433 | 3.396 | | 3.194 | | 3.311 | 3.432 |
| A-5 | | 3.076 | | 3.683 | 3.722 | 3.592 | | 3.69 | | 4.118 | 3.679 |
| Averaged Layer A | | 3.15 | | 3.405 | 3.429 | 3.369 | | 3.377 | | 3.536 | 3.381 |
| B-1 | | 3.197 | | 3.216 | 3.232 | 3.223 | | 3.234 | | 3.211 | 3.239 |
| B-2 | | 3.171 | | 3.286 | 3.262 | 3.284 | | 3.273 | | 3.27 | 3.277 |
| B-3 | | 3.199 | | 3.235 | 3.225 | 3.226 | | 3.128 | | 3.234 | 3.371 |
| B-4 | | 3.189 | | 3.196 | 3.223 | 3.205 | | 3.222 | | 3.197 | 3.307 |
| B-5 | | 3.182 | | 3.380 | 3.365 | 3.373 | | 3.386 | | 3.479 | 3.479 |
| B-6 | | 3.422 | | 3.094 | 3.098 | 3.108 | | 3.109 | | 3.123 | 3.101 |
| Averaged Layer B | | 3.227 | | 3.235 | 3.234 | 3.237 | | 3.225 | | 3.252 | 3.335 |
| C-1 | | 3.265 | | 3.230 | 3.226 | 3.225 | | 3.222 | | 3.234 | 3.385 |
| C-2 | | 3.151 | | 3.185 | 3.193 | 3.19 | | 3.202 | | 3.201 | 3.342 |
| C-3 | | 3.185 | | 3.272 | 3.264 | 3.218 | | 3.225 | | 3.205 | 3.188 |

| | | | | NH-eq | PH-eq | | PCl-eq | | PO2OH-PO-eq | PO2OH-OH-eq |
|---|---|---|---|---|---|---|---|---|---|---|
| C-4 | 3.212 | | 3.213 | 3.218 | 3.265 | | 3.257 | | 3.255 | 3.22 |
| C-5 | 3.231 | | 3.329 | 3.311 | 3.334 | | 3.328 | | 3.384 | 3.41 |
| C-6 | 3.263 | | 3.167 | 3.178 | 3.168 | | 3.169 | | 3.176 | 3.174 |
| Averaged Layer C | 3.218 | | 3.233 | 3.232 | 3.233 | | 3.234 | | 3.243 | 3.309 |
| D-1 | 3.215 | | 3.177 | 3.182 | 3.18 | | 3.184 | | 3.194 | 3.063 |
| D-2 | 3.125 | | 3.326 | 3.319 | 3.324 | | 3.325 | | 3.351 | 3.136 |
| D-3 | 3.242 | | 3.254 | 3.252 | 3.254 | | 3.25 | | 3.248 | 3.52 |
| D-4 | 3.238 | | 3.13 | 3.135 | 3.131 | | 3.132 | | 3.122 | 3.403 |
| D-6 | 3.136 | | 3.111 | 3.123 | 3.114 | | 3.12 | | 3.116 | 3.101 |
| Averaged Layer D | 3.191 | | 3.205 | 3.207 | 3.206 | | 3.207 | | 3.212 | 3.281 |
| A-B | 6.346 | | 6.681 | 6.692 | 6.542 | | 6.626 | | 7.153 | 6.643 |
| B-C | 6.344 | | 6.4321 | 6.428 | 6.424 | | 6.44 | | 6.539 | 6.538 |
| C-D | 6.301 | | 6.352 | 6.35 | 6.364 | | 6.369 | | 6.392 | 6.421 |
| | | | | **NH-eq** | **PH-eq** | | **PCl-eq** | | **PO2OH-PO-eq** | **PO2OH-OH-eq** |
| A-1(Y) | | | | 2.182 | 2.734 | | 2.796 | | 2.211 | 2.326 |
| A-2 | | | | 3.315 | 3.294 | | 3.284 | | 3.34 | 3.102 |
| A-3 | | | | 3.324 | 3.528 | | 3.404 | | 3.576 | 3.468 |
| A-4 | | | | 3.426 | 3.283 | | 3.198 | | 3.206 | 3.538 |
| A-5 | | | | 3.546 | 3.25 | | 3.323 | | 3.413 | 3.262 |
| Averaged Layer A | | | | 3.159 | 3.218 | | 3.201 | | 3.149 | 3.343 |
| B-1 | | | | 3.225 | 3.204 | | 3.204 | | 3.199 | 3.211 |
| B-2 | | | | 3.268 | 3.242 | | 3.264 | | 3.274 | 3.168 |
| B-3 | | | | 3.223 | 3.217 | | 3.226 | | 3.22 | 3.333 |
| B-4 | | | | 3.212 | 3.213 | | 3.203 | | 3.171 | 3.415 |
| B-5 | | | | 3.347 | 3.314 | | 3.311 | | 3.382 | 3.323 |
| B-6 | | | | 3.106 | 3.247 | | 3.196 | | 3.254 | 3.343 |
| Averaged Layer B | | | | 3.230 | 3.240 | | 3.234 | | 3.250 | 3.290 |
| C-1 | | | | | 3.227 | 3.229 | | 3.229 | | 3.247 | 3.277 |
| C-2 | | | | | 3.189 | 3.189 | | 3.185 | | 3.197 | 3.302 |
| C-3 | | | | | 3.216 | 3.209 | | 3.208 | | 3.188 | 3.261 |
| C-4 | | | | | 3.264 | 3.260 | | 3.264 | | 3.256 | 3.221 |
| C-5 | | | | | 3.318 | 3.303 | | 3.307 | | 3.334 | 3.291 |
| C-6 | | | | | 3.172 | 3.22 | | 3.204 | | 3.221 | 3.231 |
| Averaged Layer C | | | | | 3.231 | 3.235 | | 3.233 | | 3.241 | 3.270 |
| D-1 | | | | | 3.18 | 3.175 | | 3.174 | | 3.188 | 3.064 |
| D-2 | | | | | 3.32 | 3.319 | | 3.323 | | 3.354 | 3.089 |
| D-3 | | | | | 3.252 | 3.253 | | 3.253 | | 3.246 | 3.483 |
| D-4 | | | | | 3.132 | 3.129 | | 3.125 | | 3.106 | 3.46 |
| D-6 | | | | | 3.114 | 3.132 | | 3.123 | | 3.129 | 3.123 |
| Averaged Layer D | | | | | 3.205 | 3.207 | | 3.205 | | 3.211 | 3.274 |
| A-B | | | | | 6.477 | 6.268 | | 6.290 | | 6.542 | 6.409 |
| B-C | | | | | 6.400 | 6.429 | | 6.407 | | 6.492 | 6.428 |
| C-D | | | | | 6.347 | 6.356 | | 6.345 | | 6.366 | 6.339 |

**Table S2.** Bond angle in cationic and neutral PSK-sm system.

| | Cationic PSK-sm system | | | | | | | | | | |
|---|---|---|---|---|---|---|---|---|---|---|---|
| | prinsitine | | CN-ax | =NH-ax | NH2-ax | PH2-ax | | PCl2-ax | POMe2-ax | POOH2-PO-ax | POOH2-OH-ax |
| 2-Pb-3 | 90.069 | | 93.626 | 94.869 | 102.584 | 91.298 | 86.832 | 90.256 | 93.776 | 98.933 | 99.736 |
| 3-Pb-4 | 87.103 | | 87.715 | 86.293 | 87.446 | 84.082 | 91.754 | 84.546 | 84.734 | 88.379 | 82.901 |
| 2-Pb-5 | 86.455 | | 88.856 | 89.395 | 95.827 | 93.062 | 91.999 | 91.497 | 91.094 | 93.837 | 98.183 |
| 3-Pb-5 | 94.234 | | 97.603 | 104.357 | 109.239 | 103.966 | 115.677 | 106.262 | 107.773 | 105.476 | 107.569 |
| 4-Pb-5 | 96.433 | | 96.568 | 95.700 | 96.555 | 103.471 | 105.331 | 100.154 | 98.717 | 100.029 | 93.890 |
| 1(Y)-Pb-2 | 90.319 | | 92.167 | 87.223 | 91.627 | 77.386 | 87.985 | 81.937 | 87.550 | 78.542 | 100.930 |
| 1(Y)-Pb-3 | 174.784 | | 94.586 | 110.112 | 97.101 | 102.097 | 126.618 | 103.283 | 106.170 | 101.712 | 100.912 |
| 1(Y)-Pb-4 | 92.254 | | 82.094 | 87.167 | 70.348 | 88.025 | 77.921 | 88.906 | 83.455 | 83.927 | 65.264 |
| 1(Y)-Pb-5 | 90.982 | | 167.680 | 145.526 | 150.214 | 152.406 | 117.581 | 149.746 | 146.045 | 152.603 | 142.300 |
| | | | **CN-eq** | | **NH2-eq** | **PH2-eq** | **PMe2-eq** | **PCl2-eq** | **POMe2-eq** | **POOH2 PO-eq** | **POOH2 OH-eq** |
| 2-Pb-3 | | | 87.710 | | 96.802 | 100.195 | | 91.564 | 87.417 | 89.607 | 90.570 |
| 3-Pb-4 | | | 88.187 | | 85.453 | 82.497 | | 92.814 | 84.689 | 89.677 | 94.925 |
| 2-Pb-5 | | | 89.223 | | 94.314 | 95.161 | | 91.359 | 88.264 | 91.938 | 84.965 |
| 3-Pb-5 | | | 100.059 | | 111.285 | 109.533 | | 116.517 | 96.360 | 110.241 | 110.826 |
| 4-Pb-5 | | | 97.630 | | 96.343 | 93.697 | | 97.114 | 102.498 | 99.957 | 97.771 |
| 1(Y)-Pb-2 | | | 82.727 | | 96.538 | 88.772 | | 92.085 | 91.450 | 85.378 | 89.480 |
| 1(Y)-Pb-3 | | | 112.423 | | 112.244 | 125.766 | | 130.064 | 138.544 | 143.008 | 150.332 |
| 1(Y)-Pb-4 | | | 93.094 | | 71.304 | 81.307 | | 76.182 | 87.900 | 87.735 | 83.232 |
| 1(Y)-Pb-5 | | | 146.063 | | 133.321 | 122.869 | | 113.157 | 125.043 | 106.547 | 98.730 |
| Neutral PSK-sm system | | | | | | | | | | | |
| | | **defect** | | **=N-ax** | **NH-ax** | **PH-ax** | | **PCl-ax** | | **PO2OH-PO-ax** | **PO2OH-OH-ax** |
| 2-Pb-3 | | 94.811 | | 84.358 | 82.971 | 83.346 | | 81.694 | | 80.96 | 89.933 |
| 3-Pb-4 | | 86.974 | | 91.618 | 99.155 | 96.538 | | 97.475 | | 88.723 | 93.417 |
| 2-Pb-5 | | 86.84 | | 84.373 | 78.984 | 82.839 | | 82.323 | | 76.462 | 84.594 |
| 3-Pb-5 | | 97.584 | | 96.76 | 99.139 | 108.702 | | 108.17 | | 83.168 | 99.882 |
| 4-Pb-5 | | 93.651 | | 92.086 | 99.194 | 96.451 | | 94.112 | | 82.234 | 91.83 |
| 1(Y)-Pb-2 | | | | 92.059 | 90.572 | 89.895 | | 100.429 | | 111.412 | 94.025 |

| | | | | NH-eq | PH-eq | | PCl-eq | | PO2OH-PO-eq | PO2OH-OH-eq |
|---|---|---|---|---|---|---|---|---|---|---|
| 1(Y)-Pb-3 | | | 107.791 | 96.141 | 107.082 | | 103.897 | | 86.043 | 97.507 |
| 1(Y)-Pb-4 | | | 93.045 | 90.637 | 90.901 | | 83.722 | | 87.846 | 88.555 |
| 1(Y)-Pb-5 | | | 154.741 | 160.23 | 142.298 | | 147.86 | | 165.476 | 162.552 |
| | | | | **NH-eq** | **PH-eq** | | **PCl-eq** | | **PO2OH PO-eq** | **PO2OH-OH-eq** |
| 2-Pb-3 | | | | 79.959 | 84.592 | | 87.727 | | 78.092 | 88.704 |
| 3-Pb-4 | | | | 97.421 | 87.982 | | 98.597 | | 85.514 | 87.006 |
| 2-Pb-5 | | | | 79.663 | 86.88 | | 88.476 | | 83.8 | 86.421 |
| 3-Pb-5 | | | | 107.192 | 111.014 | | 113.471 | | 95.945 | 106.95 |
| 4-Pb-5 | | | | 101.994 | 102.382 | | 98.726 | | 94.519 | 91.977 |
| 1(Y)-Pb-2 | | | | 92.607 | 92.218 | | 99.112 | | 95.765 | 90.339 |
| 1(Y)-Pb-3 | | | | 108.989 | 133.052 | | 124.401 | | 141.348 | 150.177 |
| 1(Y)-Pb-4 | | | | 87.457 | 87.712 | | 76.94 | | 99.275 | 94.865 |
| 1(Y)-Pb-5 | | | | 140.965 | 115.566 | | 121.764 | | 121.572 | 102.734 |